# Convergent electronic-photonic systems for future XG wireless communication

**Nikhil Navaratna[1,2], Muhammad Helmi Bin Azmi[1,2], Sonu Kumar[3], Rithwik Premanand[4], G. Arun Kumar[5], A. S. Madhukumar[4], Arokiaswami Alphones[3], Aarthy Nagarajan[6], Ranjan Singh[7]***

*[1]Division of Physics and Applied Physics, School of Physical and Mathematical Sciences,*

*Nanyang Technological University, Singapore 637371*

*[2]Centre for Disruptive Photonic Technologies, The Photonics Institute, Nanyang Technological University, Singapore 639798*

*[3]School of Electrical and Electronic Engineering, Nanyang Technological University Singapore 639798*

*[4]College of Computing and Data Science, Nanyang Technological University Singapore 639798*

*[5]Department of Electronics and Communications Engineering, National Institute of Technology, Warangal, India 506004*

*[6]Department of Computer Science and Engineering, University of Notre Dame, IN, USA*

*[7]Department of Electrical Engineering, University of Notre Dame, IN, USA*

** Corresponding Author- rsingh3@nd.edu*

## Abstract

The next wireless revolution will be driven by the convergence of vast spectral resources, intelligent networks and advanced hardware platforms. As carrier frequencies advance toward the terahertz (THz) band (0.1–10 THz), future networks gain access to bandwidths capable of supporting data rates beyond 100 Gbps, while facing challenges in power generation, propagation loss, device integration, beam control and scalability. In this Review, we argue that THz communication will emerge as a convergent electronic–photonic platform that combines the compactness, integrability and programmability of electronics with the bandwidth, phase coherence and low-loss distribution capabilities of photonics. We examine the device and system foundations of this convergence, including THz sources, amplifiers, antennas, modulators, photonic interconnects, hybrid terahertz free-space-optical (TeraFSO) links and AI-enabled communication and hardware-design frameworks. By positioning THz wireless and TeraFSO systems as complementary technologies, we outline a roadmap for adaptive, ultra-high-capacity communication networks beyond 6G.

## Introduction

High speed, high fidelity and low latency communication systems have ushered humanity to unprecedented levels of progress by streamlining information transfer. The rapid rise in data rates has been facilitated by large bandwidths available at higher carrier frequencies and technologies capable of operating at such elevated frequencies. Carrier frequency scaling has culminated at the current 5G cellular technology with bands centered at the mmWave range from 24.25 GHz to 52.6 GHz featuring a 27-fold increase in carrier frequencies and approximately a million-fold increase in bandwidths from 1G technology. Substantial increases in data rates are observed when innovations in the physical layer and network layer enable spectrally efficient utilization of available bandwidths. Extrapolating the progress achieved by carrier frequency scaling over the last four decades into the future indicates that the next generations of cellular and wireless communication systems, from 6G to XG, could operate at sub-terahertz (THz) frequencies (0.1-10 THz) and beyond supporting bandwidths of several tens to hundreds of GHz, providing the potential for unprecedented data rates.

The higher possible data rates, capabilities to synergize with FSO (Free-space optical) systems and the availability of regulatory standards such as the IEEE 802.15.3d-2023[1] make THz the epicenter for the next communication revolution. The favorable positioning of THz frequencies between the electronic and photonic domains has aided the proliferation of wireless and wired (interconnect-based) devices associated with both electronic and photonic paradigms. At elevated frequencies, although electronic devices are compact and reconfigurable, they exhibit higher Ohmic losses and lower bandwidths. Conversely, photonic technologies offer broad bandwidth, low-phase-noise signal generation and seamless optical-fibre interfacing, although their THz output power, packaging and integration remain challenging. Moreover, the system requirements of future 6G–XG communications vary with the application scenario and consequently influence the choice of electronic, photonic and convergent technologies. As illustrated in Fig. 1, short-range links prioritize bandwidth and integration, whereas longer-range links impose greater demands on transmit power, antenna gain and link margin. For example, chip-to-chip and inter-rack links can benefit from low-loss photonic interconnects, indoor wireless links from high-bandwidth THz transceivers, and longer-range backhaul links from high-gain antenna arrays and, where appropriate, hybrid THz–FSO architectures**.** Photonic technologies offer advantages in bandwidth, low-noise signal generation and optical interfacing, while electronic technologies remain important for amplification, beamforming, and signal processing. These complementary capabilities motivate convergent approaches tailored to the requirements of each application.

Additionally, integrating low-loss photonic interconnects for cm-scale links with electronic circuitry enables energy-efficient signal transport while simultaneously utilizing the immense programmability offered by electronics[2]. THz-FSO (TeraFSO) schemes[3], have been proposed to achieve weather-resilient links. FSO links are highly sensitive to atmospheric effects such as fog, rain, turbulence, and scintillation, motivating these TeraFSO schemes that exploit the complementary propagation characteristics of the two frequency regimes. Therefore, THz can also bridge the gap between RF and FSO systems while research in utilizing MIMO diversity, adaptive optic systems, novel modulation schemes further enhance FSO system performance. Beyond communications, the large bandwidths, short wavelengths and high spatial resolution offered by THz frequencies also position them as a key enabler for integrated sensing and communication (ISAC), where a common hardware platform simultaneously supports wireless connectivity and environmental sensing[4,5].

In this review article, starting from progress at the electronic component level, we outline recent developments in electronics-based THz transceiver systems and emphasize how integration with state-of-the-art photonic devices can further elevate key performance metrics of THz transceivers. We describe how the scalability of THz convergent electronic-photonic systems enables communication systems (Figure 1), featuring THz and TeraFSO links, over the next generation of communications from 6G-XG.

(a)

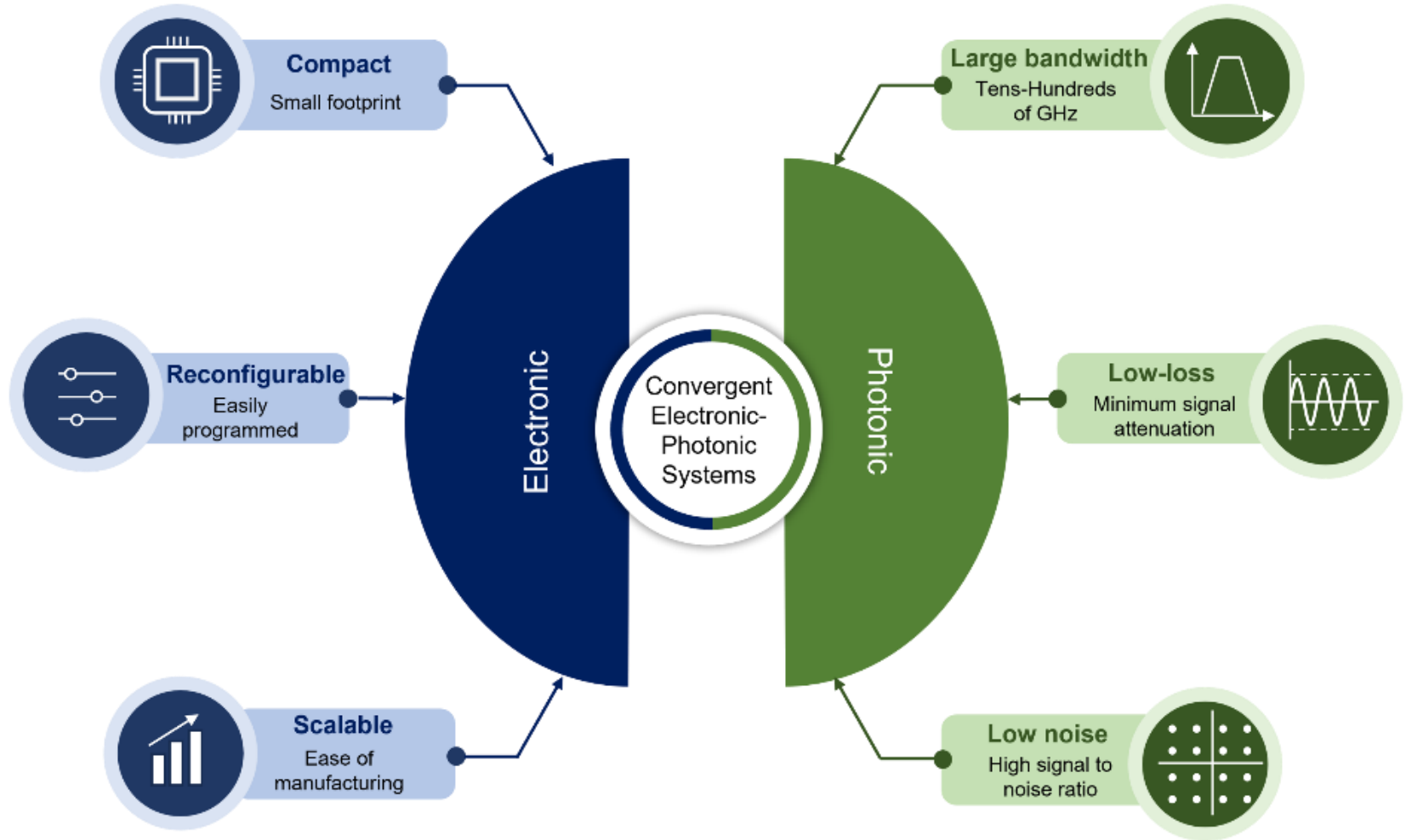

(b)

| Scenario | Target data rate | Typical range | Key challenge | How convergence helps |
|---|---|---|---|---|
| Chip-to-chip interconnects | 100 Gbps - Tbps | 1 cm – 10 cm | Interconnect loss and bandwidth-density | Low-loss photonic waveguides connecting electronic circuits |
| Indoor THz wireless | 100+ Gbps | 1 – 10 m | Limited output power | Photonic THz generation and electronic beamforming |
| Backhaul and fronthaul | Tbps and beyond | 100m – 1 km | Limited output power, atmospheric attenuation | TeraFSO networks |

**Figure 1: Vision for convergent electronic-photonic systems: (a)** Schematic illustrating how convergent systems will feature the beneficial traits of both electronics and photonics. **(b)** Table showcasing various scenarios for convergent electronic-photonic systems and how convergence helps mitigate key challenges.

## Electronic devices

Advancements in electronic circuit performance proceed in unison with improvements in the performance of constituent transistors such as: heterojunction bipolar transistors (HBT), high electron mobility transistors (HEMT) and field effect transistors (FET). However, the lack of CMOS transistors with $f_{max}$, the frequency at which unilateral power gain becomes unity, beyond 200 GHz, has hindered the development of CMOS based sub-THz electronic circuitry, driving research into processes that could overcome this bottleneck such as Silicon-Germanium (SiGe) and III-V transistors achieving $f_{max}$ 650 GHz[6] and > 1.1 THz[7] respectively. Consequently, recent progress in SiGe and III-V transistor platforms has shifted the focus from merely extending transistor cutoff frequencies to enabling complete THz transceiver architectures with sufficient power and bandwidth for future 6G-XG systems.

Various interconnected electronic components can be fabricated together to develop all-electronic THz transceiver chips. Circuit components such as signal generators, phase locked loops, power amplifiers, low-noise amplifiers and antennas play a pivotal role in THz transceivers with reliable link performance[8]. An understanding of the current state-of-the-art of these components and research trajectories can unveil avenues for further improvement and potential limitations that are difficult to surpass using all-electronic approaches. Table 1 lists the key components of THz transceivers. Target metrics towards achieving feasible link-budgets are delineated. The table and the following discussion consider these electronic technologies in the context of THz transceiver architectures, with emphasis on their achievable frequency, output power, bandwidth, and noise performance, as well as the resulting limitations for communication links.

| Component | Target metrics | | |
|---|---|---|---|
| Signal generators | $P_{out}$ (dBm) | High SNR signal that can drive subsequent stages | > 0 |
| | DC-To-RF Efficiency | Converts more of the supplied power to usable RF. Indicates low parasitics and optimized matching. | >5% |
| Power amplifiers | $P_{sat}$ (dBm) | Delivers stronger signal to antenna. A high $P_{sat}$ indicates ability to handle high output power without distortion. | > 10 |
| | PAE | Efficient power amplifier operation with low wasted DC power. | > 5% |
| Phase locked loop | Phase noise (dBc/Hz) at 1 MHz | Low phase noise signal. | < -90 |
| Antenna | Directivity (dBi) | Efficient concentration of radiated power in a desired direction. | > 15 |
| | Radiation efficiency (%) | Minimizes power lost in antenna | > 70% |
| Low noise amplifier | Gain (dB) | Amplifies weak input signal ensuring subsequent stages operate above noise floor. | > 10 |
| | Noise Figure (dB) | Low NF ensures preserves signal SNR. | < 7 |

**Table 1: Transceiver components and target metrics**. Electronic components are listed in the order they appear in all-electronic transceivers. Metrics to evaluate their performance are listed along with their description and target metrics. These metrics have been obtained using a link-budget analysis for a 100 Gbps link over a 1km distance with 30 GHz bandwidth centered at 300 GHz and a survey of the current state-of-the-art technologies.

## Signal generators and phase locked loops

On-chip signal generators are critical blocks that provide local oscillator (LO) signals required for frequency conversion and enabling the high radiated powers needed to compensate for the free-space path loss (FSPL) at THz frequencies[9]. Importantly, FSPL increases with the square of carrier frequency resulting in higher propagation losses at elevated frequencies. Signal generators eliminate the coupling and insertion losses associated with off-chip signal sources. However, at THz frequencies and beyond, generating sufficient output power while maintaining high spectral purity becomes challenging. Typical on-chip signal generator modules can either be oscillator-based[10] or frequency multiplier based[11], which have all been demonstrated in CMOS[12], SiGe[13] and InP[14] platforms. While oscillators are area and power efficient, they often exhibit lower output powers at higher frequencies than frequency multipliers. In contrast, frequency multiplier chains can generate higher carrier frequencies but are hindered by conversion losses, reduced overall efficiencies and increased power consumption. Recent advances in oscillator architecture[15], oscillator arrays[16], and frequency multiplication schemes[17,18] continue to push the attainable frequency and output power limits of all-electronic THz sources.

Resonant tunneling diodes (RTDs) featuring epitaxially grown InGaAs/AlAs on high electron mobility InP substrates are another viable alternative for on-chip signal generation with fundamental oscillations demonstrated up to 1.98 THz[19–21]. Optimizing integrable RTDs to efficiently generate power beyond 1 mW (0 dBm) for high power THz signal generation[20] remains an active area of research.

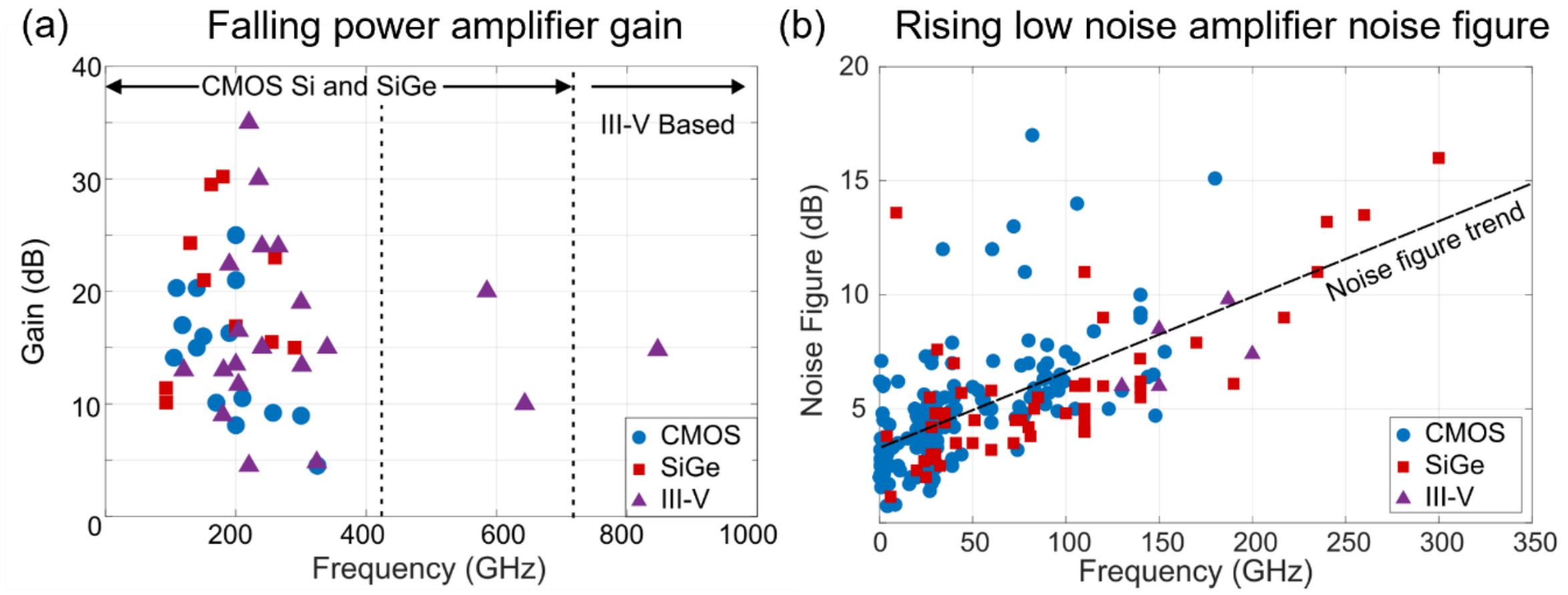

**Figure 2: Performance trends in electronic THz amplifiers**. (a) Survey of power-amplifier gain versus operating frequency across CMOS, SiGe and III–V technologies. (b) Survey of low-noise-amplifier noise figure versus operating frequency for the same technology families. Dotted trend line indicates the increase in noise figure of about 3 dB every 100 GHz. Figures 2a-b use power amplifier[22] and low noise amplifier surveys[23]

While output power determines the link budget, spectral purity is important for realizing high-capacity communication systems. LO phase noise causes frequency drift and SNR degradation, hindering attempts to achieve high data rates using LO-dependent modulation schemes such as quadrature amplitude modulation (QAM), quadrature phase shift keying (QPSK) and orthogonal frequency division multiplexing (OFDM). Consequently, improvements in signal-generation hardware must be accompanied by techniques that stabilize the generated carrier and suppress phase fluctuations.

On-chip phase locked loops (PLLs) overcome this issue by enabling LO units that can support accurate, stable and high throughput THz data-communication links[24]. On-chip PLLs at THz frequencies have been demonstrated using InP based double-heterojunction bipolar transistors (DHBT)[25], SiGe HBTs[26] and CMOS[27] processes. But achieving spectrally pure phase-locked sub-THz signals using all- electronic PLLs still remains challenging due to losses in low quality factor (Q) passives and lower supply voltages.

Despite continued progress, simultaneously achieving high output power and low phase noise remains a key bottleneck for all-electronic THz signal generators, motivating the design of photonic PLL schemes[28] with lower overall phase noise.

## Power amplifiers and low-noise amplifiers

Integrated power amplifiers (PAs) amplify THz signals for transmission across large distances. The objectives of PA design encompass high gain, bandwidth, saturated output power ($P_{sat}$) and high power-added efficiency (PAE). Ultimately, meeting these requirements is hindered by performance degradation at high frequencies close to $f_{max}$. Additionally, the increasing use of complex M-ary modulation format entails designing amplifiers with high linearity to minimize amplitude and phase distortions[29] in transmitter and receiver stages. Figure 2a illustrates the reduction in power amplifier gain with increasing carrier frequency, highlighting the growing challenges of signal amplification in transmitters operating at higher frequencies. Beyond output power and gain, bandwidth is an important consideration for THz power amplifiers. Conventional techniques to improve gain, such as positive feedback and stacking transistors, result in narrowband performance, limiting applicability to wideband systems. Future amplifiers therefore need to jointly improve output power and bandwidth while maintaining acceptable efficiency and gain.

Efforts to overcome these limitations focus on novel amplifier architectures for operation near transistor $f_{max}$ and transistor technologies offering higher $f_{max}$ and power-handling capabilities. Recent CMOS demonstrations have leveraged novel design techniques[30,31] to maintain gain and output power at frequencies near 300 GHz. Power-combining techniques[32] and novel topologies[33] with higher-$f_{max}$ have also enabled efficient sub-THz SiGe PA implementations. III-V technologies such as Gallium Nitride (GaN) HEMTs have enabled watt-level outputs up to 230 GHz[34] while InP HBT PAs[35] have even facilitated gains of 20 dB and beyond. Looking ahead, high-$f_{max}$ platforms such as SiGe BiCMOS and InP-on-Si, combined with advanced techniques such as power combining and gain boosting, could improve PA efficiency while addressing integration challenges.

Low noise amplifiers (LNAs) are among the first active stage of THz receivers and are responsible for amplifying highly attenuated incoming signals while introducing minimal additional noise. Consequently, LNA design requires both high gain and low noise figure (NF) to preserve receiver sensitivity and suppress noise contributions from subsequent receiver stages according to Friis' equation. A high performance LNA should provide an NF < 7 dB while simultaneously maintaining sufficient gain, which is challenging for CMOS, SiGe and even III-V technologies.

The primary limitation arises from increasing parasitic effects and degraded transistor performance at elevated carrier frequencies. As shown in Figure 2b, noise figure scales with rising carrier frequencies, leading to degraded receiver sensitivity and reduced link budget at higher frequencies. Recent advances in device technologies and near-$f_{max}$ design techniques[36–38] have facilitated sub-THz LNA design, although simultaneously achieving high gain and low NF remains a critical bottleneck.

Although PAs and LNAs operate at opposite ends of the transceiver, both are fundamentally constrained by transistor performance at THz frequencies. Advances in transistor technologies with higher $f_{max}$ therefore benefit both amplifier classes and remain critical for extending the range and throughput of future THz communication systems.

## Antennas and Metamaterials

Terahertz (THz) antennas play a pivotal role in radiating electromagnetic waves and addressing the limited output powers and high path losses[39]. Antenna design is driven by the need to achieve high effective isotropic radiated power (EIRP) with dynamic control. Antenna-on-chip approaches leverage the wavelength of THz-range frequencies to implement compact antennas in transceiver circuits[40–42]. Extending these approaches to silicon platforms enables designs such as λ/2-spaced antenna elements, [43] integrable silicon lenses[44] and localized backside etching[45] for high EIRP THz radiation. Arrays of sub array architectures have also been deployed to realize ultra-massive Multiple Input Multiple Output (UM-MIMO) communication[46]. The successful integration of antennas with Uni-traveling-carrier photodiode (UTC-PD)[47] and oscillators[48] across CMOS[49], SiGe[50] and III-V platforms[51] further demonstrates the feasibility of fully integrated THz transmitters. Collectively, these developments indicate that scaling antenna arrays remains a key strategy for extending communication range and improving link budgets in future THz systems. Beyond on-chip implementations, aperture antennas[52] and compact lens-based demonstrations[53] have also emerged as practical solutions.

While antenna arrays provide the radiated power and spatial multiplexing needed for long-range links, future THz networks also require dynamic control of electromagnetic wave propagation. THz metamaterials offer this additional functionality by enabling beam steering, beam focusing, and wavefront engineering[54,55]. Integrating metamaterials with antenna arrays facilitates low-power beamforming architectures with independent control over phase, amplitude, and polarization, thereby enhancing massive-MIMO capabilities[56]. Recent electromagnetic-circuit co-design approaches have demonstrated CMOS-compatible metasurfaces in which high-speed transistor switches dynamically program individual subwavelength unit cells, enabling adaptive synthesis of electromagnetic wavefronts[57]. Moreover, incorporating polarization control[58] and diversity techniques into beamforming architectures can facilitate MIMO implementations, which, alongside beam steering, scanning, and switching, will be essential for maintaining high-quality communication links and tracking in future THz communication and ISAC systems[4,59].

Looking beyond conventional CMOS implementations, emerging electronic metadevices based on high-mobility InAlN/GaN heterostructures[60] and 2D materials such as graphene[61] offer opportunities for enhanced beam manipulation at frequencies approaching and exceeding CMOS cutoff frequencies.

### From component optimization to system optimization

Strides in circuit component optimization entail detailed analysis of ideal transceiver configurations for THz communications. Recent works have showcased the efficacy of superheterodyne configurations over homodyne for low Error Vector Magnitude (EVM) THz communications across various modulation formats[62] However, currently, individual progress at circuit-component level has not yet translated to all-electronic transceiver chips that achieve high data rates at distances exceeding 10 m even at carrier frequencies beyond 250 GHz. Limitations including elevated phase noise, restricted bandwidth, metallic interconnect losses, and reduced output power continue to constrain link budgets. These challenges suggest that further progress will require not only component-level improvements but also system-level optimization. In this context, THz photonics offers a complementary pathway through low-noise signal generation, larger available bandwidths, and seamless integration with optical-fiber infrastructure. Theoretical calculations have shown link margins of 10 dB for THz photonic systems using simple intensity modulation schemes[63]. THz photonics enables the full utilization of the higher bandwidths possible at the THz frequency range and beyond by facilitating multi-carrier transmission and providing a great avenue for interfacing with optical fiber-based technologies[64].

## Photonic devices

Ideas stemming from THz photonics ranging from all-Si functionalized interconnects enabling unprecedented data rates, to low phase noise signal generators, signal converters such as photonics assisted analog-to-digital converters (ADCs), and digital-to-analog converters (DACs), can significantly mitigate link budget gaps for both wired and wireless THz propagation. Photonics based devices that can play key roles in futuristic communication systems are depicted in Figure 3. The following photonic technologies are discussed in terms of their roles in THz communication systems, including signal generation, high-bandwidth interconnects, and signal conversion, together with their demonstrated performance and remaining integration challenges.

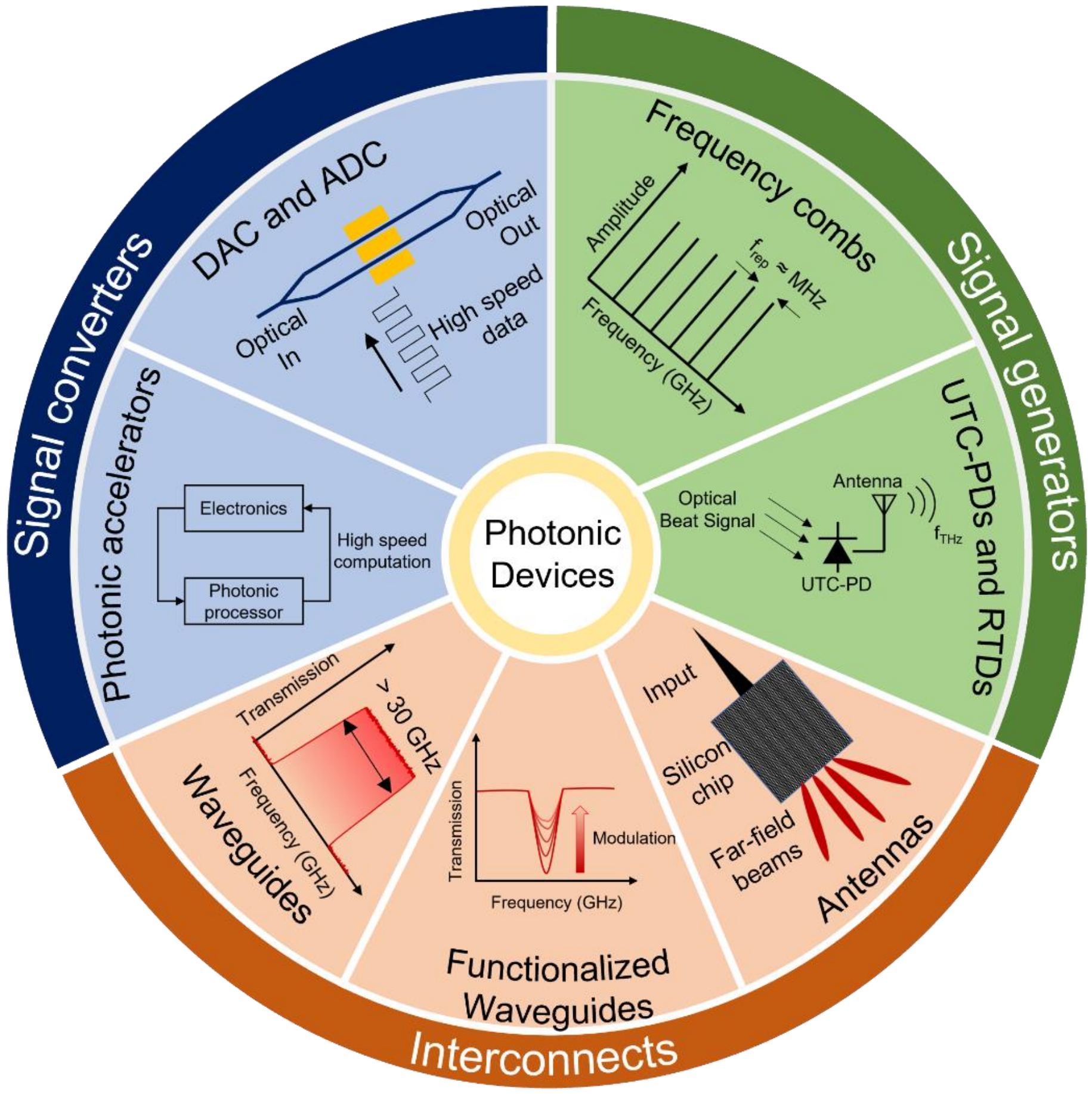

**Figure 3: Photonic devices for 6G-XG communications.** Signal generators, interconnects and signal converters are listed as various avenues in which photonic devices can feature in convergent electronic-photonic systems.

## Silicon Interconnects

Interconnects with higher bandwidth and energy density are required to keep pace with rising aggregate data rate requirements and connecting THz transceiver modules. Copper-based electrical interconnects suffer from high attenuation at THz frequencies, while optical interconnects require complex integration schemes[65]. All-Si THz dielectric waveguides offer a viable solution by capitalizing on Si's low losses at THz frequencies and facile integration with existing CMOS processes[66]. Key goals in Si waveguide design involve increasing bandwidth density, reducing group velocity dispersion with low overall energy requirements. Higher bandwidths increase the total number of available channels leading to higher aggregate data capacities. On-chip waveguides offering channel bandwidths of 30 GHz could achieve optimal energy per bit at low bit-error-rates (BERs) in interconnect-based applications[66]. Photonic crystal (PC) waveguides fabricated on high resistivity Si feature photonic band gap driven wave propagation and confinement. Bend-engineered photonic crystal waveguides have achieved losses < 0.1 dB/cm for frequencies 315 GHz – 330 GHz [67]. Further optimizing the bandwidth and bends has facilitated the design of high bandwidth spectrally efficient miniaturized waveguides that support data rates of 327 Gbps[68] across 4 spectral channels within a single waveguide. A key advantage of photonic crystal-based design is their scalability to higher frequencies with diligent optimization of the unit cell[69]. Such flexibility has resulted in waveguiding with 350 GHz bandwidths between 750 GHz – 1100 GHz[70]. Low loss transmission through multiple bends has been shown on waveguides incorporating novel physics of photonic topological insulators (PTI)[71], enabling data rates of 160 Gbps[72] with QAM-16. Further bandwidth enhancement can allow all-Si PTI platforms to support higher data rates creating bandwidth, energy, and area efficient interconnects. Effective media-clad dielectric waveguides rely on refractive-index engineering to achieve the index contrast for in-plane confinement offer another alternative to designing high bandwidth Si waveguides supporting data rates up to 30 Gbps[73].

Furthermore, a myriad of functionalities can be engineered using THz-Si waveguides to form efficient THz photonic circuitry. Ranging from simple signal routing architectures like power splitters, bandwidth splitters, Y junctions to signal manipulation and processing units such as filters can be implemented across all three waveguiding approaches. Cavities can be incorporated onto photonic crystal waveguides and topological photonic waveguides to realize switches, filters[74] and demultiplexers[72]. Active tuning can be achieved by incorporating necessary biasing units or by photoexcitation with energy greater than that of Si's bandgap[75]. Diplexers enabling 1.5 Gbps (on-off modulation) and 75 Gbps (QAM-32) have been shown on photonic crystal[76] and topological photonic platforms[77] respectively.

The guided-wave functionalities can also be coupled to free space using on-chip antennas. Si based antennas exhibit high realized gains, efficiencies, and bandwidths in on-chip antenna configurations. Photonic crystal-based designs have also facilitated the development of integrated dielectric resonator antennas[78] for endfire radiation and multibeam antenna with Luneburg lens[79] for operation at THz frequencies. External-lens free topological antennas, which utilize a circular graded index buffer to momentum match the topological edge modes to free-space, have been demonstrated to enable a 100 Gbps links over 1.2 cm[80]. Beamforming has also been demonstrated using all-Si platforms[81]. Effective-media based antennas have also shown data transmission of 4K TV signals with data rates of 6 Gbps using on-off modulation[82]. Active functionalities have been demonstrated in on-chip effective media based Si antennas where photoexcitation enables switching between ports [83]. In addition to active voltage-based control and photoexcitation, phase change materials also offer novel pathways to actively tune radiation from Si based systems. While all-electronic antennas show higher levels of tunability, broadband Si based antennas can play key roles as compact and planar alternatives to bulky off-chip antenna-lens configurations.

## Photonic signal generators

Limited output power, high phase noises and high jitter limit the scope of all-electronic THz sources in realizing low-latency network requirements which utilize complex modulation schemes[24]. Here, photonics-assisted THz sources can play a key role in advancing THz communications[64]. Photonic THz sources involve generating continuous-wave (CW) THz signals by photomixing two slightly detuned infrared laser beams in a photodiode such as a Uni-traveling-carrier photodiode (UTC-PD)[84]. Attempts to improve the overall power levels involve using high power UTC-PDs[85] and UTC-PD arrays[86]. Demonstrations of UTC-PDs at 600 GHz bands further bolster the efficacy of photonics-enabled techniques at high frequencies[87]. Besides UTC-PDs, PIN photodiodes (PIN-PDs) are also widely employed for broadband THz generation[88,89]. Recent advances in thin-film lithium niobate ($LiNbO_3$) have further expanded the landscape of photonic THz sources. These devices utilize the large second order nonlinear susceptibility of $LiNbO_3$ and enable efficient THz generation via optical rectification and difference-frequency generation, while also supporting high-speed electro-optic modulation and photonic integration[90–92].

The spectral purity of the emitted THz depends on the SNR of the optical sources and the stability between their phase and frequency. Consequently, sources with high SNR and avenues for mode-locking enable stable, spectrally pure THz generation. Spectral separation between chosen narrow-linewidth high SNR optical beams can be locked to limit the frequency drift of the mixed THz beam and reduce phase noise. Extremely low phase noises of about -95 dBc/Hz at 10 kHz offset and -110 dBc/Hz at 100 kHz offset have been achieved using optical injection locking to Kerr microcombs[28] and self-injection locking using Brillouin cavities[93] respectively. Injection-locked generation and comb-line based extraction open avenues for spectrally pure THz generation. Translating these functionalities to photonic integrated chips (PICs)[94,95] can achieve THz sources with low overall noise, mitigating the link budget gap especially for frequencies beyond 1 THz. PIC based THz transmitters featuring monolithically integrated distributed feedback lasers have already achieved data rates exceeding 100 Gbps at distances of 10 m. In addition to simplifying modulating to one of the comb lines[96] or injection-locked laser[97], THz photonic sources can realize 'optical-to-high-speed radio' interfaces featuring seamless links between THz and fiber-optic interfaces[64].

## Photonic signal converters

In convergent electronic-photonic THz systems, photonic signal converters such as ADCs and DACs serve as key enabling technologies at the electronic-photonic interface by supporting ultra-high bandwidth signal processing and mitigating electronic sampling bottlenecks. Typically operating at optical frequencies, these photonic technologies offer higher capacity, enhanced bandwidth, and lower timing jitters than traditional electronic approaches. Photonic technologies can perform the role of DACs, photonic computing accelerators and ADCs[98,99]. ADCs facilitate high bandwidth signal processing units required for communication systems operating at high carrier frequencies. Nyquist sampling theorem dictates that the ADC sampling rate must be at least twice that of the bandwidth of the data signal. However, ADC sampling rates have struggled to keep pace with increasing bandwidths [100]. Photonic assisted ADCs can overcome the timing jitter-bottleneck to achieve higher sampling rates without trading off the resolution[101]. Typically, sampling involves multiplying the signal with a train of mode locked laser (MLL) pulses in a non-linear element, such as a Mach-Zehnder modulator (MZM) [101]. While the attosecond jitter of MLL pulses enables much higher sampling rates than purely electronic counterparts, the devices are complex, bulky, and not easily integrable. A potential solution can be optical sampling based on convoluting the signal spectrum with a rectangular frequency comb [102]. Such source-free all optical sampling has also been demonstrated on silicon photonic MZM [103] showing great promise for facilitating sampling rates of 300 GSa/s and 400 GSa/s. In addition, concurrently developing schemes for photonic assisted digital-to-analog converters can further augment sampling rates and efficiencies[104]. Photonic based ADC and DAC methods, especially electronic-photonic integrated chips (EPICs), show immense signal processing capabilities for higher generations of communication technology due to their enhanced bandwidths, higher sampling rates, low jitter, resilience to noise and direct interoperability with fiber-optics.

THz wireless schemes have been demonstrated throughout the sub-THz range up to 625 GHz[105]. Moreover, as summarized in Table 2, several photonic THz wireless demonstrations meet IEEE 802.15.3d-2023 requirements of operating channels between 252 GHz and 450 GHz. However, all-photonic THz systems are not without their issues. Additional improvements are required for output power and thermal management of UTC-PDs. Furthermore, bulky off-the-shelf photonic modules such as lasers, optical modulators, amplifiers, complicate packaging. Developing photonic integrated circuits (PICs) especially on monolithic InP, reduce the size, weight, power and packaging costs across a wide range of frequencies[106,107] up to several THz. Key characteristics of electronic transceivers such as modulation-complexity, reconfigurability and active tunability are still difficult to achieve on all-photonic systems. Optimizing various methods of integrating electronics and photonics can unlock these immense benefits.

| Type | Technology | Frequency (GHz) | Data rate (Gbps) | Distance (m) | Data rate × Distance (Gbps ·m) | External antenna and/or lens gain(dBi) | Modulation format | IEEE 802.15.3d Requirements | Reference |
|---|---|---|---|---|---|---|---|---|---|
| Electronic | InP HEMT 35 nm | 240 | 64 | 850 | 54,400 | 55.1 | QPSK | Not met | 110 |
| | InP HEMT 35 nm | 300 | 56 | 10 | 560 | 38 | QAM 16 | Met | 111 |
| | SiGe HBT 130nm | 240 | 100 | 1 | 100 | 25 | QPSK | Not met | 112 |
| | MMIC Front-end | 300 | 44.8 | 1000 | 44800 | 51 | QAM 4 | Met | 108 |
| | MMIC Front-end | 300 | 12.6 | 645 | 8127 | 62 | QAM 32 | Met | 109 |
| | SiGe HBT 130nm | 240 | 20 | 0.15 | 3 | 14 | BPSK | Not met | 45 |
| | CMOS 40nm | 265.68 | 80 | 0.03 | 2.4 | 24 | QAM 16 | Met | 113 |
| | CMOS 40nm | 260 | 10 | 0.04 | 0.4 | +5 dBm (EIRP) | OOK | Met | 114 |
| Photonic | Kerr Microcomb | 300 | 60 | 0.5 | 30 | 25 | QAM 16 | Met | 28 |
| | Stimulated Brillouin Scattering lasers | 300 | 205.6 | 214 | 43998 | 26 | QAM-32 | Met | 115 |
| | Photonic source | 320 | 50 | 850 | 42500 | 48.5 | QAM-16 | Met | 116 |
| | Monolithic PIC source | 408 | 131 | 10.7 | 1401 | 51 | QAM-16 | Met | 117 |
| | DFB Laser source | 200 | 144 | 0.47 | 67.68 | 26 | QAM-16 | Not met | 118 |
| Convergent Electronic - Photonic | Photonic source MMIC receiver | 237.5 | 100 | 20 | 2000 | >50 | QAM 16 | Not met | 96 |
| | Photonic source, vacuum-electronic amplifier | 335 | 27.84 | 2200 | 61248 | 58 | QAM 16 | Met | 119 |

**Table 2:** Summary of data communication experiments whose innovations are largely in electronic transceivers[45,108–114] , which emphasize photonic front ends or receivers[28,115–118] and those that push convergent electronic-photonic[96,119] approaches featuring novelties in both electronic and photonic components.

## Integration schemes

At a chip-level, methods of integrating electronic and photonic elements include monolithic, heterogeneous, and hybrid integration[3]. Monolithic integration involves fabricating all components on the same substrate, whereas hybrid integration relies on subsequently connecting separately fabricated photonic and electronic chips. Heterogeneous integration involves integrating technologies at the wafer or die level. Monolithic integration is common at lower frequencies where active and passive devices can be implemented on the same semiconductor platform. Monolithic integration is particularly suited for compact high-speed electronic front-end components such as amplifiers, mixers, phase-locked loops, and antenna arrays where minimizing parasitics and interconnect losses is critical. However, at THz frequencies, restricting designs to a single substrate limits the range of available electronic and photonic technologies, while large differences in feature sizes and isolation requirements further complicate implementation. In contrast, hybrid integration facilitates platform diversity and allows photonic and electronic elements to be independently characterized and optimized before integration[120] (Figure 4).

Consequently, hybrid integration approaches have facilitated most convergent THz electronic-photonic demonstrations. UTC-PD transmitters and RTD-based receivers have been integrated on broadband THz dielectric waveguides[121–123]. Bonded THz interconnects[124,125] have been demonstrated where dielectric waveguides were seamlessly integrated with electronic circuitry. Availability of lead-free, solder-free and low-temperature bonding technologies is a major catalyst for adopting bonded THz systems. Similarly, silicon-interposer-based substrate-integrated waveguide filters operating at 280 GHz and 420 GHz have demonstrated the feasibility of interposer technologies for THz system[126]. Figure 4 highlights the key photonic and electronic devices that are potentially suited for future integrated convergent systems.

Beyond hybrid integration, heterogeneous approaches such as micro-transfer printing offer a pathway towards integration of THz electronic and photonic technologies[127]. By transferring devices fabricated on their native substrates onto CMOS-compatible platforms, micro-transfer printing enables integration of advanced III-V electronic and photonic components. Such approaches aid InP-on-Si integration, combining the high-frequency performance of InP amplifiers and oscillators with silicon-based devices. Additionally, photonic components such as lasers and modulators can be integrated with conventional electronics platforms to realize advanced heterogeneous systems[128].

The multitude of integrable systems with scope for further optimization and co-integration make THz frequencies the genesis for achieving communication systems from 6G to XG.

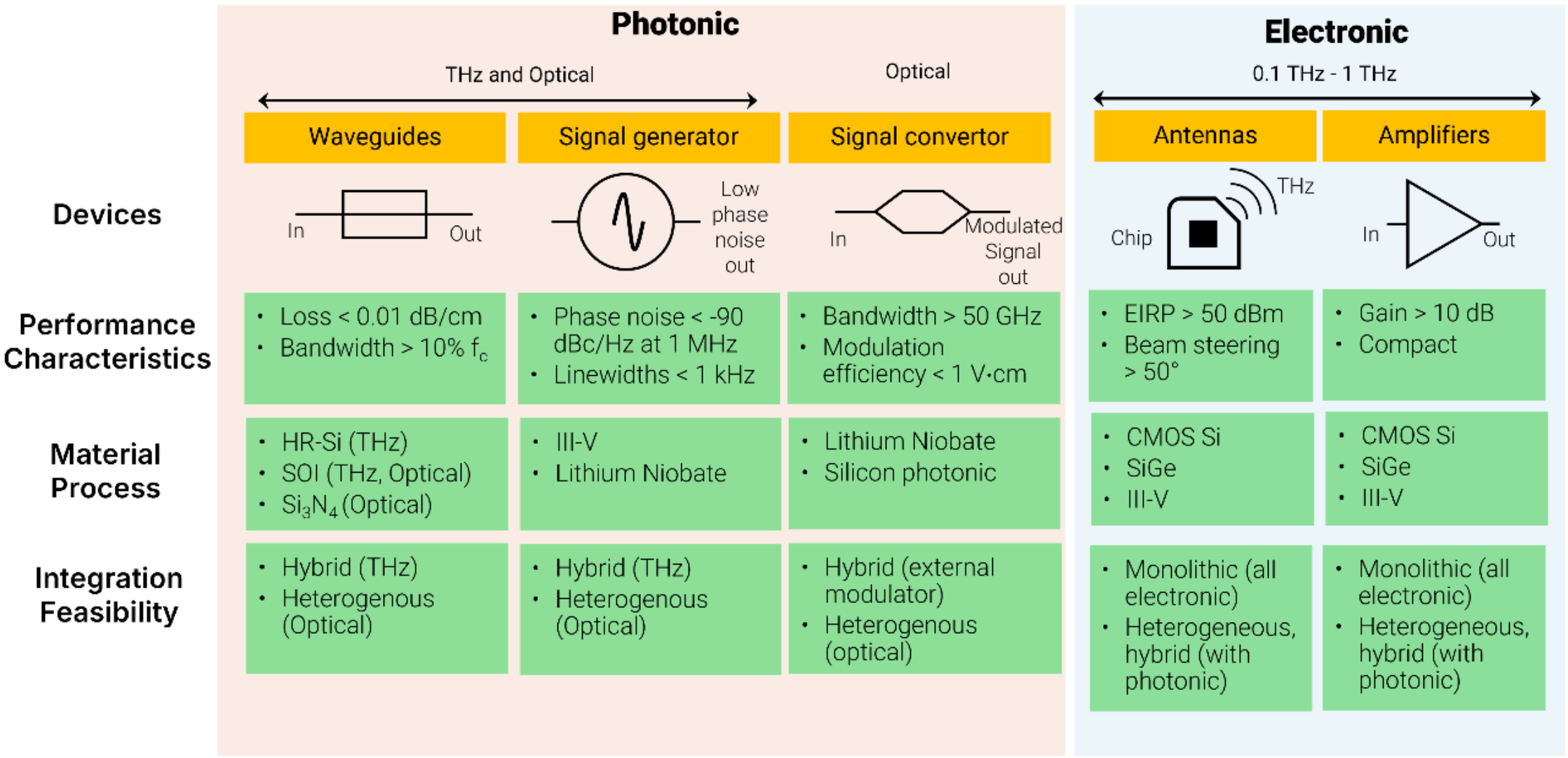

**Figure 4: Potential constituents of convergent photonic-electronic circuits**: 6G to XG communication networks will require harmonizing photonic components and electronic components. The various material processes and schemes for on-chip integration are highlighted.

## Outlook

The transition from 5G to emerging 6G-XG systems of the future will require more than incremental improvements in device performance. As illustrated in Figure 5, communication systems have historically exhibited an approximately $10^{X-4}$ Gbps increase in data rates across successive generations. While this trend serves as a heuristic reference rather than a quantitative forecast, future communication architectures operating at THz frequencies and beyond will need to effectively harness the vast bandwidths available at higher frequencies. In contrast to earlier generations constrained by the lack of bandwidths, high frequency links are limited by atmospheric attenuation and molecular absorptions. The most promising path forward is the convergence of electronic and photonic technologies across transceivers and network architectures. Three research directions appear particularly critical for realizing this vision.

Firstly, the primary hardware challenge for future communication systems is improving link budgets without sacrificing energy efficiency. Addressing this challenge will require advances in integrated electronic-photonic hardware. In photonics, future efforts should include photonic interconnects with lower dispersion to facilitate high spectral efficiency and low energy/bit signal transfer between electronic circuits[129–131]. Future photonic THz sources will require simultaneous improvements in output power, phase stability, energy efficiency and chip-scale integration to translate their bandwidth advantages into practical convergent transceivers. In parallel, innovations in circuit architectures and design methodologies that extend the practical operating frequencies of CMOS technologies, alongside advancements in high-$f_{\max}$ SiGe and III–V technology nodes, will enable high frequency transceivers with improved output power and communication ranges. Success in these areas could enable compact, energy-efficient transceivers capable of Tbps scale wireless and interconnect applications.

Secondly, the growing complexity of convergent electronic-photonic systems necessitates AI-assisted approaches in both hardware design and network operation. AI-assisted inverse design is also expected to accelerate the development of electronic and photonic components by exploring design spaces beyond traditional template-driven approaches[132]. Beyond component design, AI can support adaptive waveform generation, beam management, resource allocation and real-time network orchestration across highly dynamic environments[133,134]. A major challenge is the lack of AI models that accurately capture the underlying physics of THz electronic and photonic devices[135] while remaining computationally efficient[136]. Many existing models are also too resource-intensive for deployment on energy-constrained edge platforms. Research efforts should therefore focus on compact physics-informed models, federated learning approaches and neurosymbolic frameworks[137] that combine data-driven learning with physical and logical constraints. Together, these developments could improve the performance, adaptability, and operational efficiency of future THz networks.

Thirdly, realizing the cumulative benefits of carrier-frequency scaling deeper beyond the THz frequencies, advanced electronic-photonic hardware, and AI-driven optimization will require resilient network architectures capable of operating under dynamic environmental conditions. Here, TeraFSO systems represent a particularly promising direction. Rather than viewing THz and FSO technologies as competing solutions, future networks can exploit their complementary characteristics to maximize capacity, availability and resilience, particularly as standalone FSO links remain susceptible to weather-induced attenuation, turbulence effects and practical deployment constraints[138,139]. For instance, TeraFSO schemes could involve FSO femtocells to bolster short-range MIMO links via wavelength-division multiplexing based scaling while relying on THz systems for long-range links (> 1 km). Further research is needed to develop intelligent link-selection and switching protocols, joint resource-management strategies and integrated backhaul architectures that dynamically allocate traffic between THz and optical wireless links under varying environmental conditions. As communication systems progress towards higher frequencies, their physical dimensions also evolve, with shorter wavelengths enabling more compact antennas and dense integration, while propagation loss and atmospheric attenuation increasingly constrain practical link distances. These trade-offs motivate different system architectures and integration approaches across 6G–XG applications. The convergence of communication and sensing functionalities is expected to further enhance these architectures, enabling integrated sensing and communication (ISAC) systems that simultaneously support high-capacity data transfer, localization, imaging, and environmental awareness.

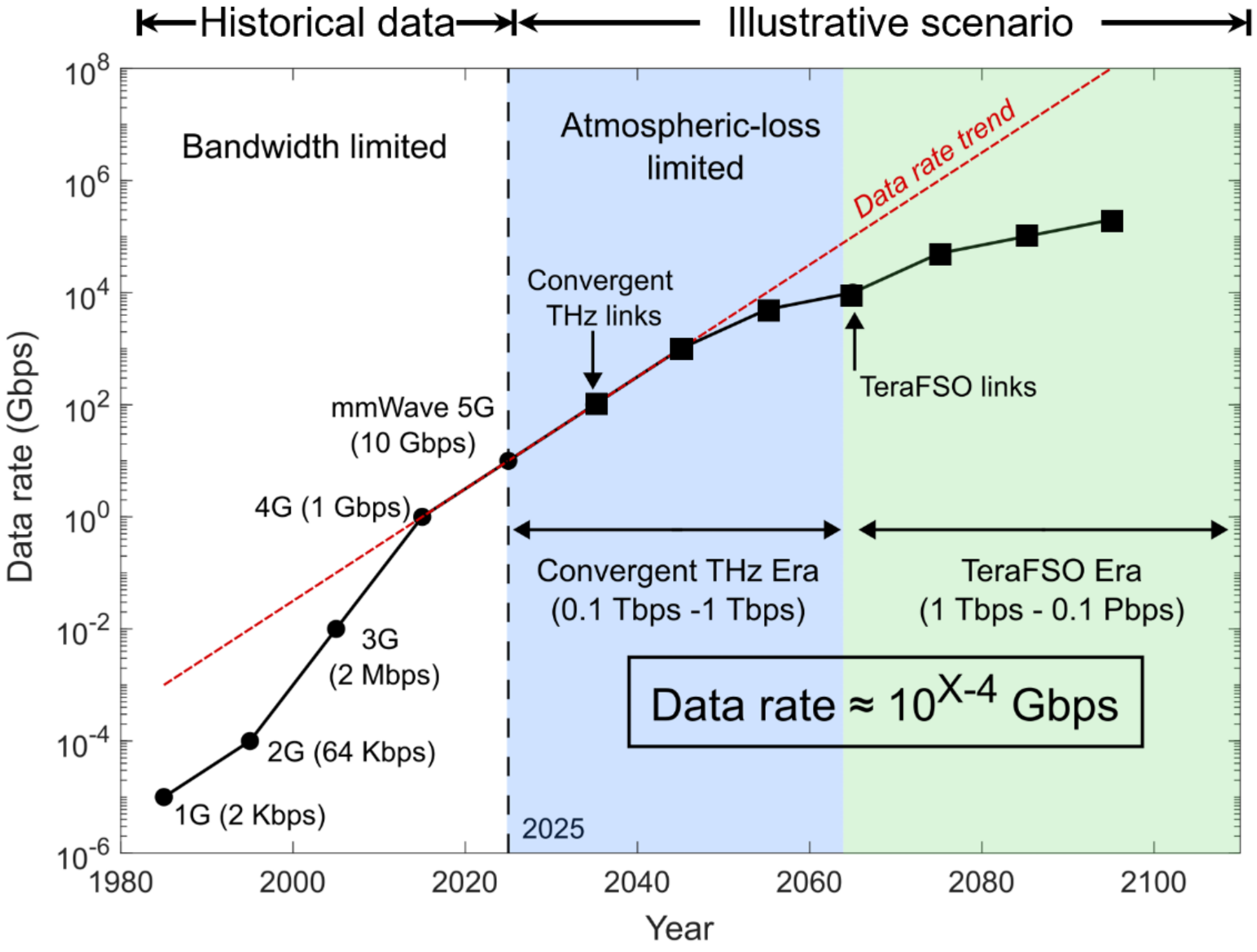

**Figure 5: Illustrative roadmap for data-rate progression with year.** (a) Data rate (Gbps) plotted against year. The red dotted line indicates the idealized trend in which data rates scale with generation number “X” as $10^{X-4}$ Gbps. The blue shaded region indicates the convergent THz era, in which advances in convergent THz technologies may support higher data rates, while the green shaded region indicates the TeraFSO era, in which the complementary use of THz and FSO systems may enable substantially higher aggregate data rates. Circular markers denote historical data, whereas square markers denote illustrative future scenarios. The projected values, including those in the Pbps regime, represent illustrative potential aggregate capacities rather than forecasts of single-link data rates.

Together, these developments outline a pathway towards communication systems capable of sustaining the continued growth in network capacity envisioned in Figure 5. By combining advanced electronic-photonic hardware, AI-native optimization, and resilient TeraFSO network architectures, future 6G-XG systems could transform the immense bandwidths available at THz and optical frequencies into practical, scalable, and energy-efficient communication infrastructures for an increasingly connected world.

## Acknowledgements

All the authors acknowledge the research funding support from National Research Foundation (NRF) Singapore, Grant No: NRF-CRP23-2019-0005 and NRF-MSG-2023-0002.

## Author Contributions

N.N., R.S. conceptualized the review paper. N.N., M.H.B.A. performed the literature review and contributed the initial draft. K.S., R.P. wrote the sections on antenna and free space optical communication respectively. A.N. wrote the section on AI accelerated future communications. A.A., A.S.M., G.A.K. provided feedback and suggestions. N.N, M.H.B.A, R.P., A.N., and R.S. wrote the manuscript with input from all the other co-authors. R.S. led the overall project.

## Competing interests

The authors declare no competing interests.

**Key Points**:

- Terahertz (THz) frequencies provide tens to hundreds of GHz of bandwidth, creating a pathway toward wireless links with data rates approaching 100 Gb/s and ultimately beyond.
- Electronic THz transceivers are constrained by declining amplifier gain, increasing noise figure, resulting in persistent link-budget challenges.
- Photonic technologies offer low-phase-noise signal generation, high-bandwidth signal processing, and low-loss interconnects that can mitigate key limitations of all-electronic THz systems.
- Convergent electronic-photonic architectures combine the programmability and compactness of electronics with the bandwidth, power handling, and low-loss transmission capabilities of photonics.
- THz–free-space-optical (TeraFSO) systems exploit the complementary propagation characteristics of THz and optical links to improve communication resilience under varying atmospheric conditions.
- The convergence of THz electronics, photonics, TeraFSO networking, and AI-assisted design establishes a roadmap toward future 6G–XG infrastructures capable of aggregate throughputs approaching the petabit-per-second regime.